\documentclass[conference]{IEEEtran}
\IEEEoverridecommandlockouts

\usepackage{cite}
\usepackage{amsmath,amssymb,amsfonts}
\usepackage{textcomp}
\usepackage{xcolor}
\usepackage[normalem]{ulem}

\newcommand{\NTT}{\operatorname{NTT}}
\newcommand{\stride}{\operatorname{L}}
\newcommand{\diag}{\operatorname{D}}
\newcommand{\tensor}{\otimes}

\newcommand{\one}{\operatorname{I}}

\newcommand{\rotation}{\operatorname{R}}

\usepackage{subfigure}
\usepackage{algorithm}
\usepackage{graphicx}
\usepackage{algpseudocode}
\usepackage{multirow}
\usepackage{multicol}
\usepackage{wrapfig}
\usepackage{tabularx}
\usepackage[frozencache,cachedir=.]{minted}
\usepackage{listings}
\usepackage{hyperref}

\def\BibTeX{{\rm B\kern-.05em{\sc i\kern-.025em b}\kern-.08em
    T\kern-.1667em\lower.7ex\hbox{E}\kern-.125emX}}
\begin{document}

\title{RPU: The Ring Processing Unit}

\author{Deepraj Soni\textsuperscript{$\dagger$}\textsuperscript{$1$}, Negar Neda\textsuperscript{$\dagger$}\textsuperscript{$1$}, Naifeng Zhang\textsuperscript{$2$}, Benedict Reynwar\textsuperscript{$3$}, Homer Gamil\textsuperscript{$1$}, \\ Benjamin Heyman\textsuperscript{$1$}, Mohammed Nabeel\textsuperscript{$1$}, Ahmad Al Badawi\textsuperscript{$4$}, Yuriy Polyakov\textsuperscript{$4$}, Kellie Canida\textsuperscript{$3$}, \\ Massoud Pedram\textsuperscript{$5$}, Michail Maniatakos\textsuperscript{$1$}, David Bruce Cousins\textsuperscript{$4$}, 
Franz Franchetti\textsuperscript{$2$}, Matthew French\textsuperscript{$3$}, \\ Andrew Schmidt\textsuperscript{$3$}, Brandon Reagen\textsuperscript{$1$}

\thanks{\textsuperscript{$1$}NYU}
\thanks{\textsuperscript{$2$}Carnegie Mellon University}
\thanks{\textsuperscript{$3$}USC Information Sciences Institute}
\thanks{\textsuperscript{$4$}Duality Technologies}
\thanks{\textsuperscript{$5$}USC Viterbi School of Engineering}
\thanks{\textsuperscript{$\dagger$}Deepraj Soni and Negar Neda contributed equally to this work.}
\thanks{{Correspondence to: Deepraj Soni and Negar Neda (dss545, negar)@nyu.edu}}
}

\maketitle

\begin{abstract}
Ring-Learning-with-Errors (RLWE) has emerged as the foundation of many important techniques for improving security and privacy,
including homomorphic encryption and post-quantum cryptography.
While promising, these techniques have received limited use due to their extreme overheads of running on general-purpose machines.
In this paper, we present a novel vector Instruction Set Architecture (ISA) and microarchitecture for accelerating the ring-based computations of RLWE.
The ISA, named B512, is developed to meet the needs of ring processing workloads while balancing high-performance and general-purpose programming support.
Having an ISA rather than fixed hardware facilitates continued software improvement post-fabrication and the ability to support the evolving workloads.
We then propose the ring processing unit (RPU), a high-performance, modular implementation of B512.
The RPU has native large word modular arithmetic support, capabilities for very wide parallel processing, and a large capacity high-bandwidth scratchpad to meet the needs of ring processing.
We address the challenges of programming the RPU using a newly developed SPIRAL backend.
A configurable simulator is built to characterize design tradeoffs and quantify performance.
The best performing design was implemented in RTL and used to validate simulator performance.
In addition to our characterization, we show that a RPU using 20.5mm$^{2}$ of GF 12nm
can provide a speedup of 1485$\times$ over a CPU running a 64k, 128-bit NTT, a core RLWE workload.
\end{abstract}

\section{\bf Introduction}
The continued increase in security threats and demand for better data privacy protections has incited many to re-think security primitives and how data are computed on. From the security viewpoint, the advent of quantum computing has launched efforts into the next generation post-quantum cryptography ~\cite{computer_security_division_post-quantum_2017}.
These new encryption techniques provide security guarantees against quantum computers that break many existing schemes.
At the same time, users are demanding more control over who can see and use their personal data.
Homomorphic encryption (HE) is another emerging type of encryption where secured (encrypted) data can be computed on directly, preserving confidentiality while granting users access to high-quality services.
While each of the above techniques provides different security and privacy capabilities, they are all Ring-Learning-With-Errors (RLWE)-based algorithms with the same fundamental underlying structures: rings.
\par From a systems perspective, rings are large arrays of elements in a field (an integer and modulus).
RLWE-based security and privacy techniques have seen relatively little practical use due to their poor performance on existing hardware.
There are three major sources of slowdown that can be understood by looking at HE, an application based on RLWE. 
First, computations in HE involve modular arithmetic with large moduli.
Modular arithmetic is significantly more involved than standard integer math and is typically not supported by existing hardware, especially when large primes are used.
Second, encryption incurs a large ciphertext expansion factor.
Plaintext data must first be organized into an array and is then encrypted into two ciphertext arrays with elements ranging from dozens to thousands of bits. 
Prior work has reported up to a $50\times$ size increase after encryption \cite{feldmann_f1_2021}.
Finally, HE operations require an exorbitant amount of computation.
This is because many HE operations, e.g., HE multiplication, are not simple translations of their plaintext equivalent.
These operations require extra functions, such as the Number Theoretic Transform (NTT),
which greatly increase the number of operations needing to be processed. 
These overheads result in high latency, energy consumption, and memory pressure,
which limits the deployment of HE and ring applications.
\par To address the emerging needs of ring processing, we present the Ring Processing Unit (RPU) and B512 vector ISA.
B512 was designed to address the needs of ring processing while being programmable, as algorithms are still rapidly evolving, and to support continued software improvements post fabrication.
It has a vector length of 512 elements to increase the work per instruction while providing flexibility to the programmer, as the smallest size ring is typically one to two thousand elements.
It includes a large, local scratchpad to (double) buffer vector data, 64 vector registers, and native support for large word modular arithmetic.
The ISA was designed with simplicity in mind and has only 17 instructions to keep front-end overheads at bay.
\par The RPU is a highly parallel and decoupled implementation of B512.
Major components of the design include independent pipelines for data access, compute, and shuffle, as well as a front-end.
Vector data memory (VDM) is parameterized to support different SRAM implementation choices and memory partitioning schemes to tradeoff bandwidth, frequency, and area.
Being a vector ISA, compute can be parallelized across lanes.
We develop a novel compute lane named the HPLE, High-Performance LAW (Large Arithmetic Word) Engine, to rapidly process compute instructions.
Each HPLE is equipped with a slice of the Vector Register File (VRF) and modular arithmetic units.
Finally, to reuse data in the register file, a shuffle pipeline is used to support register-register data movement and breaking vectors.
Each pipeline is decoupled to improve resource utilization and instruction throughput.
The goal is to dedicate as much chip area as possible to the backend of the RPU and limit the overheads associated with general-purpose processing.
As efficiency demands simplicity, the front-end is intentionally kept as simple as possible.
It is in-order and does not support renaming.
We introduce a lightweight technique named busyboarding to track and block all register data dependencies while dispatching ready instructions to available pipelines. The busyboard guarantees correctness while allowing instructions to execute and complete out of order using the decoupled pipelines.
\par The challenge with having a simple front-end is high-performance programming.
RPUs provide high-performance potential, but the rigid nature of the front-end places a heavy burden on the compiler/programmer to schedule code and layout data properly.
Whenever the busyboard detects a dependence, new instructions cannot issue until they resolve.
To improve performance, programs should be scheduled around it.
We address and automate this process by developing a new SPIRAL~\cite{franchetti2018spiral} backend for the RPU.
SPIRAL has a rich library of transformations and optimizations we leverage to compile and optimize code for the RPU. 
This is especially true for NTT, which is the ring equivalent of the FFT.
As we will demonstrate, SPIRAL is able to effectively select an NTT algorithm and generate high-performance B512 programs to improve performance over a baseline program by $1.8\times$.
\par We evaluate and characterize the RPU using a detailed cycle-level simulator.
The simulator was parameterized to consider a range of different IP, namely modular multiplier design, number of HPLEs, and VDM partitioning (banking).
This enables rapid design space exploration to quantify design decisions.
Our exploration indicates that an RPU design with 128 VDM banks and 128 HPLEs maximizes performance per area when processing NTTs.
To validate our results, we implemented this design point in RTL and emulated it on a Palladium system~\cite{noauthor_palladium_nodate}.
Additional experiments are provided to quantify the benefit of optimized SPIRAL programs, the impact of multiplier design on RPU performance per area, speedup over a CPU across a range of ring sizes, and area analysis.
\par This paper makes the following contributions:
\begin{enumerate}
\item We introduce B512, a novel vector ISA tailored to the needs of ring processing.
It supports a vector length of 512 for highly-parallel execution and ample registers and local memory space to meet the memory needs of ring processing.
\item We develop the Ring Processing Unit (RPU) that implements B512 and addresses the three barriers in ring processing.
It includes decoupled pipelines to mask data movement cost, an efficient front-end to minimize general-purpose overheads, and a highly-parallel backend constituting many HPLEs to handle the heavy processing load.
\item We develop a SPIRAL backend to target RPU programming to effectively and automatically generate high-performance B512 programs.
An in-depth evaluation is done using NTT with a wide range of ring sizes.
\item We implement a cycle-level simulator to rigorously characterize the design and understand tradeoffs via design space exploration.
The simulator is validated against a RTL implementation and achieves 97\% performance accuracy. 
Our analysis shows that the best configuration is with 128 VDM banks and HPLEs.
This RPU can process a 64k, 128-bit NTT in 6.7us using 20.5mm$^{2}$ of area in GF 12nm, providing a 1485$\times$ speedup over a CPU.
\end{enumerate}
\section{\bf Preliminaries}
\subsection{\bf{Ring Learning with Error (RLWE)}}
As the amount of information in the digital domain is increasing rapidly, data security is a major concern of our time. While modern cryptography has developed solutions to protect data storage and transfer, large-scale quantum computing has the potential to break the security of existing digital infrastructure \cite{shor1999polynomial}. Against the attacks from quantum computers, Fully Homomorphic Encryption (FHE) protects data-in-use and Post-Quantum Cryptography (PQC) guarantees the security of the data, at-rest and in-motion. 
FHE schemes (such as BGV\cite{brakerski_leveled_2012}, CKKS \cite{cheon2017homomorphic}, and many more) and PQC schemes (such as CRYSTALS-Kyber \cite{bos_crystals_2018}) are RLWE-based schemes that operate on polynomial rings. RLWE is a variant of standard LWE with smaller computation complexity and a similar security guarantee \cite{lyubashevsky_ideal_2010}. The RLWE-based cryptographic algorithms perform arithmetic operations over ring $\mathbb{Z}_{q}/(x^{n} + 1)$ where $n$ is power-of-two integer representing the polynomial degree and $q$ is the prime modulus. 
\par HE, an RLWE-based cryptography algorithm, operates on rings with large polynomial degrees and modulus sizes. CKKS \cite{cheon2017homomorphic} supports operations on approximate numbers. BGV \cite{brakerski_leveled_2012}  operates on integers in the finite field. These schemes implement different applications that offer secure computation, secure storage, secure data transfer, and secure machine learning tasks \cite{bossuat_efficient_2021, dathathri2020eva, dathathri_chet_2019,gilad-bachrach_cryptonets_2016, han_logistic_2019,lee_privacy-preserving_2022,podschwadt_classification_2020}. These applications need a large modulus. Hence, we decompose the large modulus to smaller ones using Residue Number System (RNS) for faster performance.
\subsection{\bf Residue Number System (RNS)}
In RNS, a large integer, modulus $Q$, is represented by its value modulo $L$ pairwise co-prime integers, $q_{i}$, following the Chinese Remainder Theorem (CRT), where $Q = q_{0}q_{1}q_{2}...q_{(L-1)}$. These pairwise co-primes are called residue polynomials.
RNS effectively breaks each polynomial into several polynomials with smaller coefficients, achieving faster modular arithmetic implementation.
 \begin{figure}[t!]
  \centering
 \includegraphics[width=0.5\textwidth]{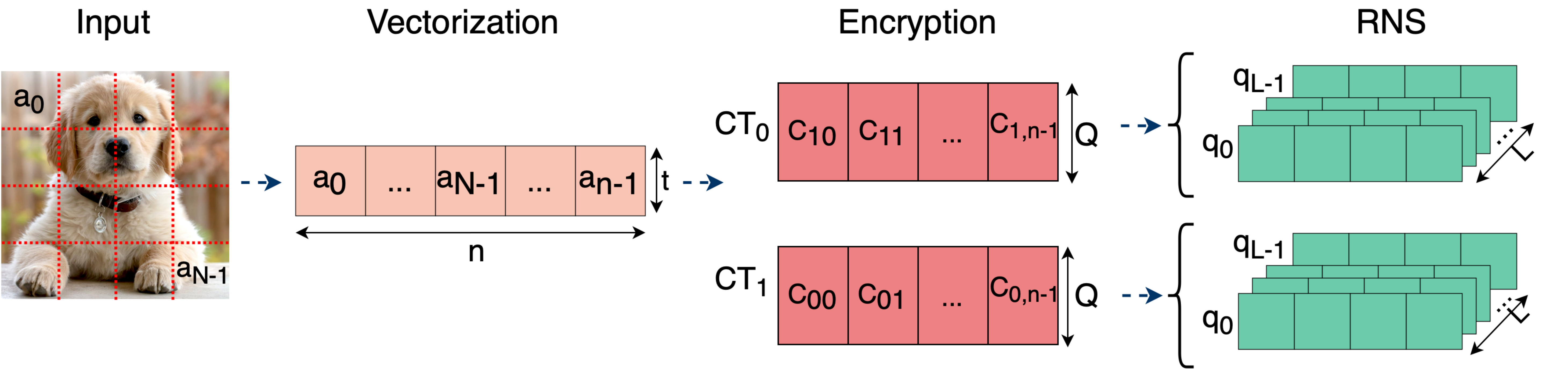}
 \vspace{-1em}
  \caption{\bf Overview of converting an input image to ciphertext and breaking down to small vectors using RNS.}
 \label{fig:rns}
   \vspace{-1em}
 \end{figure}
\par Figure~\ref{fig:rns} shows a generic example of converting an input image to ciphertext using RNS. First, we convert the input into a vectorized format with plaintext modulus $t$. Next, we encrypt the data using a public key that converts it into two ciphertext polynomials with ciphertext modulus $Q,$ where $Q$ $>> t$. As the $Q$ is quite large, we divide it into smaller moduli. In the case of ciphertexts, each pair of smaller polynomials are called \emph{towers}.
During polynomial multiplication, each tower operates independently. RNS supports a wide range of polynomial coefficient widths. If the coefficient width is larger than the modulus arithmetic, we can break it into multiple towers. For example, if we design a 128-bit modular arithmetic processor, a polynomial with 1,600-bit modulus is converted to 13 towers where each tower has 128-bit elements. Hence, we can use RNS for larger bit-widths and we can choose non-RNS computation for lower bit-widths.
\subsection{\bf NTT}
Polynomial multiplication is the process of multiplying two polynomials together to create a new polynomial.
Polynomial multiplication is a major bottleneck for efficient ring processing. Done naively, it has quadratic complexity $\mathcal{O}(n^{2})$.
These computations are typically sped up by transforming the polynomial using the Number Theoretic Transform (NTT)\cite{bos_crystals_2018}. 
NTT is the generalized form of the Discrete Fourier Transform (DFT), and it is an integral part of the RLWE algorithms for accelerating polynomial multiplications over finite fields (e.g., NTT counts for 94\% of homomorphic multiplication's execution time \cite{fan2022tensorfhe}). NTT is able to transform polynomials into a domain that reduces the complexity of polynomial multiplications to linear $\mathcal{O}(nlog\ n)$, making polynomial multiplication much faster.
\section{\bf The B512 ISA}
\label{sec:ISA}
We have designed a new vector ISA, named B512, to balance the needs of high-performance ring processing with programmable, general hardware.
An ISA was chosen over a fixed hardware design to support a wide range of potential ring applications. While algorithms are still evolving, having an ISA gives us the flexibility to map new algorithms to previously designed hardware.
The ISA has 17 64-bit instructions with spare encoding space for expansion in future work.
Instructions provide native modular arithmetic support with special consideration for commonly used computations (e.g., butterfly), various load/store patterns, and register-register vector breaking via shuffling.
B512 supports a maximum VDM size of
32 MiB and 16 MiB scalar data memory (SDM).
Unique register files (amount) are specified for vector data (64), scalar data (64), addresses (64), and moduli (64). The vector length is 512.
This was chosen to increase the work per instruction and reduce the front-end cost,
while still providing flexibility to the compiler to optimize schedules and data layout.
Rings tend to be large, e.g., the HE standard~\cite{HomomorphicEncryptionSecurityStandard} specifies a minimum ring size of 1024, and having a vector size less than the ring size provides flexibility to the compiler.
\par Table~\ref{tab:isa} reports the encoding for B512 instructions.
The instructions interact with data memories, register files, and HPLEs. 
Instructions are classified into three types: 
(1) Load/Store Instructions (LSI), 
(2) Compute Instructions (CI), and 
(3) Shuffle Instructions (SI). 
We note that in the microarchitecture, there are three decoupled pipelines that support parallel execution of each instruction type.
\begin{table}[t]
\caption{B512 architecture. Including compute, load/store, and shuffle instructions.}
\setlength{\tabcolsep}{2.8pt}
\begin{tabular}{|ccccccccc|}
\hline
\multicolumn{1}{|c|}{{[}63:55{]}} & \multicolumn{1}{|c|}{{[}54:49{]}} & \multicolumn{1}{|c|}{{[}48{]}} & \multicolumn{1}{|c|}{{[}47:44{]}} & \multicolumn{1}{c|}{{[}43:24{]}} & \multicolumn{1}{c|}{{[}23:18{]}} & \multicolumn{1}{c|}{{[}17:12{]}} & \multicolumn{1}{c|}{{[}11:6{]}} & {[}5:0{]} \\ \hline
\multicolumn{9}{|c|}{\bf Load/Store Instructions (LSI)}\\ \hline
\multicolumn{1}{|c|}{-}           & \multicolumn{1}{c|}{-}           & \multicolumn{1}{c|}{-}           & \multicolumn{1}{|c|}{Opcode}      & \multicolumn{1}{c|}{Address}      & \multicolumn{1}{c|}{VD}          & \multicolumn{1}{c|}{Mode}        & \multicolumn{1}{c|}{Value}      & RM        \\ \hline
\multicolumn{1}{|c|}{-}           & \multicolumn{1}{c|}{-}           & \multicolumn{1}{c|}{-}           & \multicolumn{1}{|c|}{Opcode}      & \multicolumn{1}{c|}{Address}     & \multicolumn{1}{c|}{-}           & \multicolumn{1}{c|}{-}           & \multicolumn{1}{c|}{RT}         & -         \\ \hline
\multicolumn{9}{|c|}{\bf Compute Instructions (CI)}\\ \hline
\multicolumn{1}{|c|}{VD1}      & \multicolumn{1}{|c|}{VT1}      & \multicolumn{1}{|c|}{BFLY}      & \multicolumn{1}{|c|}{Opcode}      & \multicolumn{1}{c|}{-}           & \multicolumn{1}{c|}{VD}          & \multicolumn{1}{c|}{VS}          & \multicolumn{1}{c|}{VT}         & RM        \\ \hline
\multicolumn{1}{|c|}{-}           & \multicolumn{1}{c|}{-}           & \multicolumn{1}{c|}{-}           & \multicolumn{1}{|c|}{Opcode}      & \multicolumn{1}{c|}{-}           & \multicolumn{1}{c|}{VD}          & \multicolumn{1}{c|}{VS}          & \multicolumn{1}{c|}{RT}         & RM        \\ \hline
\multicolumn{9}{|c|}{\bf Shuffle Instructions (SI)}\\ \hline
\multicolumn{1}{|c|}{-}           & \multicolumn{1}{c|}{-}           & \multicolumn{1}{c|}{-}           & \multicolumn{1}{|c|}{Opcode}      & \multicolumn{1}{c|}{-}           & \multicolumn{1}{c|}{VD}          & \multicolumn{1}{c|}{VS}          & \multicolumn{1}{c|}{VT}         & -         \\ \hline
\end{tabular}
\vspace{-1em}
\label{tab:isa}
\end{table}
\par \textbf{Load/Store Instruction (LSI):}
LSIs interact with data memories (VDM and SDM) and register files (Vector Register File (VRF) and Scalar Register File (SRF)). The VDM is used to store the actual rings and twiddle factors. 
Vector load and store interact with VDM and VRF to load vectors with 512 elements from the VDM to VRF and store back the results to the VDM. 
In Table~\ref{tab:isa}, RM points to a register in the Address Register File (ARF). The address of the first element in the VDM would be defined as ARF[RM] + OFFSET, and the rest of the elements are addressed based on the two other fields, MODE and VALUE. 
MODE and VALUE together implement four different addressing modes in vector load/store instructions, where the two STRIDED-SKIP and REPEATED modes enable efficient NTT implementation by transferring each $2^{VALUE}$ and skipping other $2^{VALUE}$. 
For scalar data, only a load instruction is required to load data from SDM to a register in SRF.  
\par \textbf{Compute Instructions (CI):}
B512 supports vector-vector and vector-scalar modular addition, subtraction, and multiplication. 
These instructions perform point-wise modular computation between two vectors or a vector and a single scalar value. 
There is a special butterfly instruction that combines these three modular vector-vector instructions and is used to accelerate NTT and reduce code size.
As shown in Table~\ref{tab:isa}, RM points to a modulus register to define the modulus of the computation. 
VS and VT act as input registers for modular arithmetic operations, and VD is the output register. 
VT1 and VD1 are used as the two additional registers for butterfly.
\par \textbf{Shuffle Instructions (SI):}
To improve NTT performance, B512 includes register-register instructions to merge elements of two vector registers into one. There are four different shuffle instructions:
(1) Unpack Low (UNPKLO), (2) Unpack High (UNPKHI), (3) Pack Low (PKLO), and (4) Pack High (PKHI). 
Shuffle instructions have two source registers, VS and VT and one destination register, VD. 
UNPKLO stores the first half of VS and VT elements to VD in an interleaved fashion. 
UNPKHI stores the second half of VS and VT elements to VD in an interleaved fashion. 
PKLO stores the even indexed VS elements to the first half of VD and the even indexed elements of VT to the second half of VD. 
PKHI stores the odd indexed VS elements to the first half of VD and the odd indexed elements of VT to the second half of VD. 
These instructions allow architectural vectors to be split up within the VRF, and they were chosen to support the development of efficient NTT programs leveraging SPIRAL and taking pressure off the VDM. B512 shuffle instructions are similar to those in x86~\cite{intel_shuffle}.
\subsection{\bf Parameter Selection}
\label{subsec:sel_param}
Vector width defines how many elements are processed in one instruction. HE workloads are repetitive. Hence, we prefer a larger vector width to increase the amount of parallel work per instruction. However, a large vector width will restrict the software from generating efficient code for smaller polynomial sizes and using the compute resources efficiently. We choose a vector width of 512, which is a good trade-off between parallel computation and software flexibility. RPU backend enables parallelism by transferring and computing data in parallel.
\par Each memory size is based on the requirements of the key RLWE workload, NTT. 
The Instruction Memory (IM) is 512KB and the VDM supports storing at least one complete instance of data for the 64K NTT workload, but we commonly double buffer.  
The current design operates on 128b modulus size, denoted as $|q|$, and 128b data types. With $|q|$ equal to 128b, we have the flexibility to run different HE applications with or without RNS, as reported in the homomorphic standard \cite{HomomorphicEncryptionSecurityStandard}. 
\par The register file should be small enough to directly access HPLEs with minimal latency and it should be large enough to store more values for flexibility and for minimal data movement between data memory and the register file. Hence, we choose to have 64 registers in all register files.
\section{\bf The Ring Processing Unit}
\label{sec:arch}
\begin{figure*}[t!]
  \centering
  \includegraphics[width=.95\textwidth]{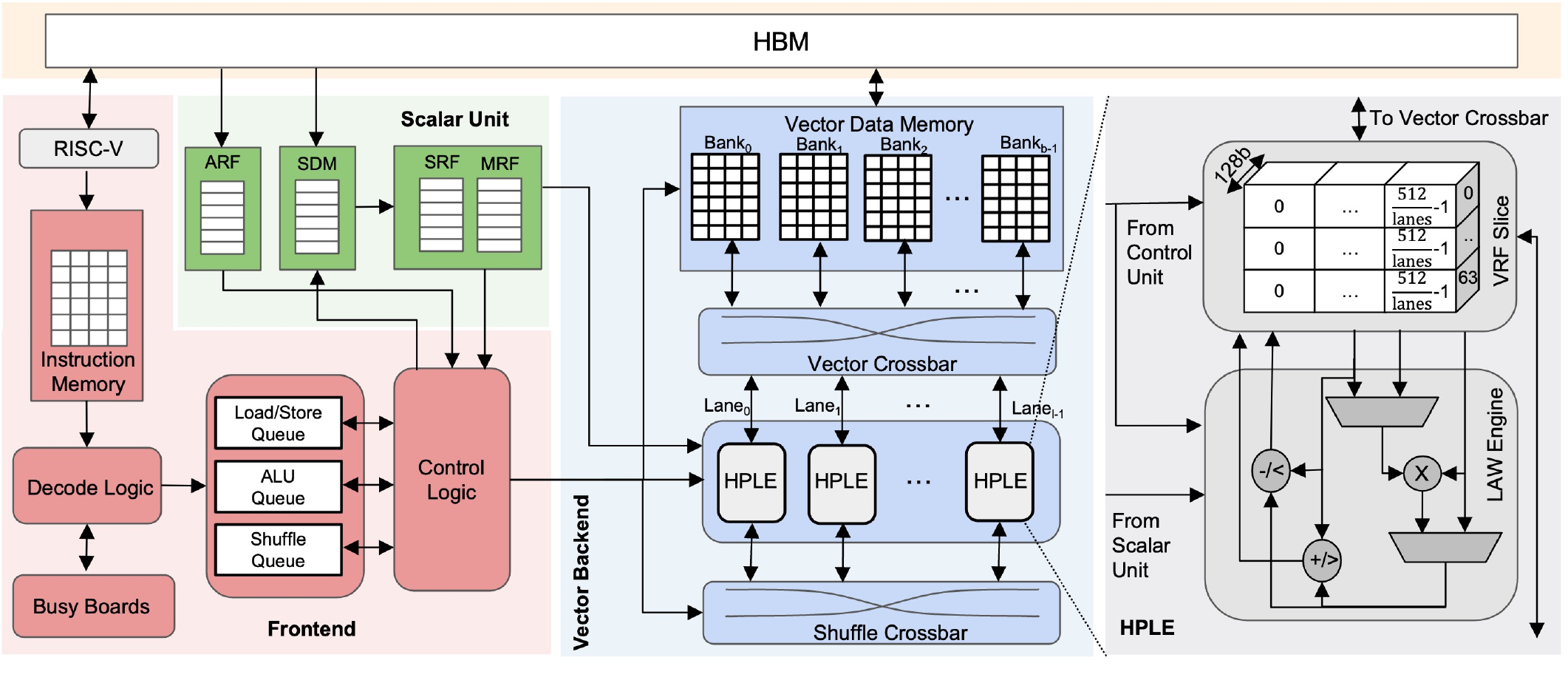}
  \vspace{-1em}
  \caption{\bf Ring Processing Unit (RPU) architecture.}
  \label{fig:tile}
  \vspace{-1.5em}
\end{figure*}

In this section, we describe the microarchitecture of the designed Ring Processing Unit (RPU). 
The RPU was designed for general ring processing with high performance by taking advantage of regularity and data parallelism. 
We achieve this balance by designing explicitly managed hardware to elide the high costs and complexity of caches, dynamic scheduling logic, and prediction, and task the compiler with handling scheduling and data movement at compile time.
As we will show, this philosophy works very well for ring processing.
Figure~\ref{fig:tile} shows an overview of the RPU. 
Based on the data parallel nature of ring processing workloads, and notably HE, parallel vector architectures are highly amenable for meeting the performance needs. Below, we cover each major component of the design.
\subsection{\bf Frontend} 
All RPU programs are stored in the local instruction memory. When a task is to be executed, a controlling RISC-V core issues a start command to the front-end with an instruction memory pointer to the first instruction of the kernel.
To mitigate the frontend overheads, the RPU is kept efficient using in-order logic and lightweight dependence tracking.
The front-end fetches and decodes instructions in-order. Data hazards are checked using a \emph{busyboard}, which is used to describe our lightweight scoreboarding technique.
The busyboard is a bit array that tracks all the vector registers being used by all in-flight instructions.
No renaming is supported, and whenever a decoded instruction register is busy, the entire front-end stalls. The design prioritizes efficiency and the area overheads are negligible. Design is also highly sensitive to instruction scheduling.
\par Once instructions clear all data hazards, they are dispatched to one of three decoupled queues:
(1) Load/store  Queue for LSIs, (2) Compute Queue for CIs, and (3) Shuffle Queue for SIs.
Once an instruction is in the queue, it can run in parallel with any other instruction as we know (from the busy board) there are no dependencies.
The parallel execution via the decoupled pipelines is a key to achieving high performance with general-purpose processing, as it masks much of the data movement time. 
\subsection{\bf RPU Backend}
The RPU backend provides the high-performance structures needed for effective ring processing.
The major components include three decoupled pipelines for
computing via High-Performance LAW Engines (HPLEs), register-register data shuffling, and Vector Data Memory (VDM).
It also includes a Scalar Data Memory (SDM) to house the constants needed by HE.
\subsubsection{\bf High Performance LAW Engine (HPLE)}
The HPLE is the computational unit in the RPU.
Each has a Long Arithmetic Word (LAW) engine and a partition of the VRF, or a VRF slice. 
The LAW Engine contains a modular multiplier, a modular adder, a modular subtractor, and two comparator units. 
NTT/iNTT is a key kernel in RLWE, and the HPLEs support native
butterfly computation via a butterfly instruction. Each CI interacts with the VRF slice and LAW Engine to perform three tasks: 
read data from the VRF slice to the LAW engine,
start the computation in LAW Engine, and
store the output to the VRF slice. 
Here we use 128b to meet the needs of HE precision, see Section~\ref{subsec:sel_param}.
The RPU allocates multiple HPLEs as lanes in classic vector designs.
Instantiating many HPLEs addresses the high volume of computation in the ring processing workloads.
\par In each HPLE, the LAW Engine is connected to the VRF slice. VRF slice is part of VRF that is divided among HPLEs. According to B512 ISA, the VRF has 64 vector registers with 512 elements.
Each slice has $64*\frac{512}{num\_HPLEs}$ elements.
If we store each register of VRF in a different memory, it will require a smaller and more efficient memory. To increase area efficiency, we stack four registers in one memory. The four registers in one memory cannot be accessed simultaneously. Therefore, instructions require special scheduling and data placement in the VRF, which is handled by SPIRAL. Hence, each VRF slice constitutes 16 single-port memories with $\frac{512*4}{num\_HPLEs}$ words.  
\par A VRF slice interacts with HPLEs, Vector Crossbar (VBAR), and Shuffle Crossbar (SBAR). 
Though the VRF has 16 SRAM ports, 10 are exposed to the three pipelines; five ports (three read ports and two write ports) for HPLEs, three ports (two read ports and one write port) for SBAR, and two ports (one read port and one write port) for VBAR. 
For computation, each VRF slice simultaneously sends the data from input registers to the corresponding HPLE. HPLE performs the computation. Once the HPLE outputs the result, the VRF slice stores it back to output registers. 
\subsubsection{\bf Shuffle Crossbar (SBAR)}
The SBAR transfers the data across VRF registers, executing SIs.
It facilitates efficient implementations of complex access patterns to improve NTT's efficiency by allowing register-register data shuffle.
With the SBAR, vectors can be broken in the VRF, saving round trips through the VDM to restructure data in B512 vectors. The SBAR supports all four modes of shuffle transfer.
\subsubsection{\bf Vector Data Memory (VDM)}
We instantiate an RPU with a 4MiB VDM.
Here, we find 4MiB is sufficient to double buffer off-chip data loading (via HBM) with the execution of a kernel. However, if more capacity is needed, the VDM can be expanded to up to 32MiB. The large word size and capacity of the VDM necessitates the use of large SRAM macros that tend to run at relatively low frequency. Therefore, the VDM limits the frequency of the RPU, as we assume a single clock domain. When the VDM has 32 banks, it operates at 1.29GHz. With 128 and 256 banks, it runs at 1.68GHz.
\subsubsection{\bf Vector Crossbar (VBAR)}
The VBAR transfers the data between the VDM and VRF slices of HPLEs and executes LSIs.
It supports all four modes of data transfer, as described in Section~\ref{sec:ISA}. HPLEs can efficiently read or write data from different VDM banks in parallel by utilizing the VBAR, which transfers the data concurrently.
In practice, we find striding data across banks resolves nearly all bank collisions. We designed a parameterized VBAR to support any number of banks and HPLEs.
\subsubsection{\bf Scalar Unit}
A Scalar Data Memory (SDM) and Scalar Register File (SRF) are included to handle the many constants needed in RLWE processing.
SDM is 32KB and uses 128b words, which loads data into SRF. The SRF sends values to HPLEs when the RPU executes scalar instruction. To add flexibility to operations, a Modulus Register File (MRF) has been added to the scalar unit.
The MRF enables modulus changing at the instruction granularity, enabling the potential to process different towers simultaneously.
SRF and MRF data are directly transferred to HPLEs. 
To enhance the flexibility, we added an Address Register File (ARF) to the scalar unit for indirect memory access. This allows moving the location of stored data in the VDM, without the need of changing instructions.

\section{\bf RPU Programming with SPIRAL}
\label{sec:SPIRAL}
As noted earlier, due to the light-weight front-end of the RPU, the performance of our design relies heavily on the sequence of instructions. To overcome this challenge, we use SPIRAL~\cite{franchetti2018spiral} for instruction generation. SPIRAL~\cite{franchetti2018spiral} is a program generation/synthesis system that takes in high-level specifications and produces highly optimized implementations. SPIRAL has generated code whose performance is comparable to expertly hand-tuned
code across kernels and platforms, especially linear transforms such as DFTs.
\par We extended SPIRAL and developed a formal framework to
capture computational algorithms (NTT) and computing platforms (RPU), using a unifying mathematical formalism we call SPIRAL's Operator Language\cite{franchetti2009operator},\cite{zhangtowards}. Specifically, we built on prior work on modular FFT (ModFFT) in SPIRAL\cite{meng2010spiral} and created the NTTX package to isolate and encapsulate all NTT-related components. While the current prototype supports the RPU and ANSI C, the code generation backend targets all \mbox{SPIRAL-supported} platforms.
 NTTX provides FFTW-like\cite{frigo1998fftw} C/C++ API in line with FFTX-like\cite{franchetti2018fftx} code generation.
 \par We defined several APIs from the host processor (C/C++ API) to the kernel (\textit{kernel code} in Listing \ref{listing:ntt-1024}) on RPU, modeled after CUDA. Launch code is used to abstract low-level system built-in constructs to convert the host C-based data structures into scratchpad-based data structures.
\par To take advantage of the large vector instructions offered by RPU, we added both Korn-Lambiotte FFT algorithm\cite{korn1979computing} and the Pease FFT algorithm\cite{pease1968adaptation} as breakdown rules to SPIRAL. The long vector length required a re-formulation of the algorithms to map the dataflow to RPU capabilities. Using the Korn-Lambiotte algorithm, NTTs of size $r^{k}$ are represented in SPIRAL's Operator Language (OL)\cite{franchetti2009operator} as \cite{zhangtowards}
\begin{equation*}
    \NTT_{r^k} = \rotation^{r^k}_r \left( \prod^{k-1}_{i=0} \stride^{r^k}_{r^{k-1}} \diag^{r^k}_i (\NTT_r \tensor \one_{r^{k-1}}) \right)
  \label{eq:ol}
\end{equation*}.
\par We implemented a functional simulator in C++ to verify the generated C code using OpenFHE\cite{OpenFHE} inputs and outputs, and an analyzer simulating RPU to count the cycles as the metric for our optimization. 
To optimize the performance of the SPIRAL-generated code on RPU while abiding by the hardware constraints, we implemented and automated register allocation, register spilling, and store-to-load forwarding for the kernel code. 
To interleave independent instructions within the kernel and thereby hide latencies, we decomposed the NTT kernel into multiple ``rectangles'' that consume parts of the input sizes in one stage, and go as deep as they can until more inputs from the starting stage are needed. At last, we used a greedy instruction scheduler to detect any easily-achieved low-level optimization, further reducing the overall cycle count. 
\par Using SPIRAL, we generated both forward and inverse vectorized radix-2 NTTs with sizes from 1,024 to 65,536, whose correctness is verified against OpenFHE data. Listing \ref{listing:ntt-1024} shows the radix-2 1,024-point NTT kernel code generated by SPIRAL. 

% \SetupFloatingEnvironment{listing}{placement=tbp}
\begin{listing}
\begin{minted}[frame=none,
        obeytabs=true,
        tabsize=4,
        % linenos,
        % numbersep=-6pt,
        escapeinside=||,
        fontsize=\scriptsize]{c}
// SPIRAL generated NTT Code for RPU vector architecture
#include <rpu.h>
// kernel code
void _ntt1024x512_b1() {
    enter(OP_DEFAULT);
    _vload_512x128i(REG_V60, REG_A1, 0);
    _vload_512x128i(REG_V20, REG_A1, 8192);
    _vbroadcast_512x128i(REG_V19, REG_A3, 1, 1);
    _vimulmod_512x128i(REG_V59, REG_V20, REG_V19, REG_M1);
    _vaddmod_512x128i(REG_V58, REG_V60, REG_V59, REG_M1);
    _vsubmod_512x128i(REG_V57, REG_V60, REG_V59, REG_M1);
    _vunpacklo_512x128i(REG_V56, REG_V58, REG_V57);
    ...
    _vstores_512x128i(REG_A2, 16, REG_V21, 2);
    leave(OP_DEFAULT);
}
\end{minted}
\caption{SPIRAL-generated radix-2 1,024-point NTT}
\label{listing:ntt-1024}
\end{listing}
\section{\bf Evaluation \& Characterization}
\label{sec:eval}
In this section, we evaluate and characterize the RPU via design space exploration to understand area-performance tradeoffs and energy consumption.
We use large NTTs in the evaluation, as NTT is the key kernel of RLWE. The analysis includes how performance scales with ring size, speedup over a CPU, and the effectiveness of code optimizations.
\subsection{\bf Methodology}
A cycle-level performance and functional simulator have been developed to study the RPU and conduct design space exploration.
The simulator, written in C++, models all aspects of the design and faithfully process B512 codes generated by SPIRAL.
The simulator is configurable to consider different VDM banking strategies, allocations of HPLEs, pipeline depths, and component IP (e.g., multiplier).
Each parameterized component (HPLE, VRF, VDM, SBAR, VBAR, SDM, SRF, MRF, and ARF) is synthesized using Synopsys' Design Compiler (DC) with GF 12nm targeting 2GHz. 
We used a commercial SRAM compiler to model all memories.
Logic units' area and frequency numbers are taken from DC. 
We report the memory power from the datasheet and other components using Synopsys DC.
32 core 2.5GHz AMD EPYC 7502.
\par \textbf{Validation:} 
We used test vectors from OpenFHE\cite{OpenFHE} for functional validation of B512 code and the RPU. 
All codes generated by SPIRAL run through the functional simulator and match OpenFHE's output.
We further implemented a full RPU design in RTL, synthesized it, and sent to a Palladium\cite{noauthor_palladium_nodate} for full system emulation. 
Our emulations also used the real inputs from OpenFHE and execute SPIRAL B512 programs under full circuit emulation. We confirmed the outputs of emulation matched outputs from openFHE. 
Our simulation performance estimates further matched within 97\%.
\subsection{\bf NTT Performance}
\label{subsec:ntt_sweep}
\begin{figure}[t]
  \centering
  \includegraphics[width=.95\linewidth, scale=0.5]{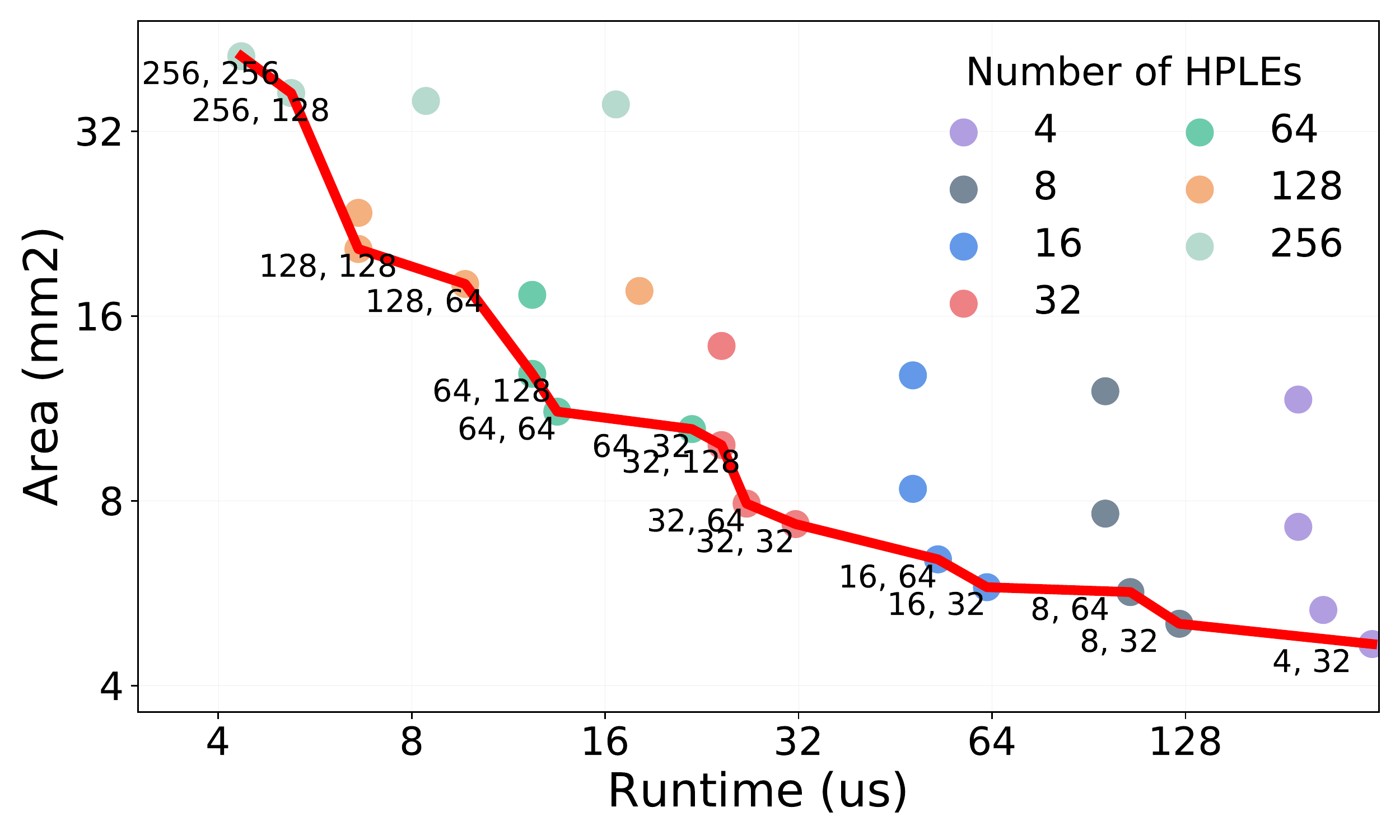}
  \vspace{-1em}
  \caption{\bf 64K NTT area-latency trade-off varying HPLEs and VDM banks. 
  Pareto optimal designs are marked as: (HPLEs, banks).}
  \label{fig:ntt_area_latency}
\end{figure}
In this section, we evaluate RPU's performance using NTT, a key kernel in all RLWE schemes.
Figure~\ref{fig:ntt_area_latency} shows the trade-off between area and runtime when running optimized NTT code on RPU, by varying the number of HPLEs and the number of VDM banks. The red line shows the Pareto optimal points. 
We denote each Pareto configuration as (HPLEs, banks). 
For example, configuration with 64 HPLEs and 128 banks are denoted as (64, 128). 
The (4, 32) configuration has the lowest area (and performance) because there are only 4 HPLEs, which require high area, and 32 VDM banks. In addition, it has the fewest connections between the VDM, HPLEs, and VRF slices, minimizing VBAR and SBAR overhead. Due to the limited parallelism, the (4, 32) configuration has the longest runtime.
The (256, 256) configuration has the highest number of parallel hardware components, maximizing performance.
The RPU frequency is set to the VDM frequency, as we find VDM runs slowest.
We run the RPU at 1.29 GHz, 1.53 GHz, 1.68 GHz, and 1.68 GHz for 32 banks, 64 banks, 128 banks, and 256 banks, respectively. 
\par To analyze RPU trends, we have examined the area and runtime for RPU configurations with both minimum and maximum resources, i.e., 4  and 256 HPLEs. 
With four HPLEs, we observe a significant increase in area, but  little improvement in runtime when doubling the VDM banks. 
The (4, 256) configuration requires 2.5$\times$ more area and has 0.75$\times$ less runtime compared to (4, 32) RPU configuration. This is because having more banks increases the load/store bandwidth, but the limited number of HPLEs is unable to process the loaded data, which limits the overall performance improvement.
With 256 HPLEs, we observe a minor area increment and a significant improvement in runtime by doubling the number of VDM banks. 
The (256, 256) configuration requires a 20\% area increment and 3.5$\times$ runtime improvement compared to the (256, 32) RPU configuration. As mentioned above, more banks create faster execution of load/store operations, which were a bottleneck for RPU configuration with 256 HPLEs. Additionally, we have found that most Pareto points correspond to the configurations where the number of HPLEs equals the number of banks or twice the number of banks. This is because such configurations provide an opportunity for continuous parallel execution for LIs and CIs; without CIs waiting for LIs to complete or vice versa.  This kind of pipeline creates fewer stalls, uses the hardware components efficiently, and provides optimized performance.
\begin{figure}[t!]
  \centering
  \includegraphics[width=0.9\linewidth, scale=0.5]{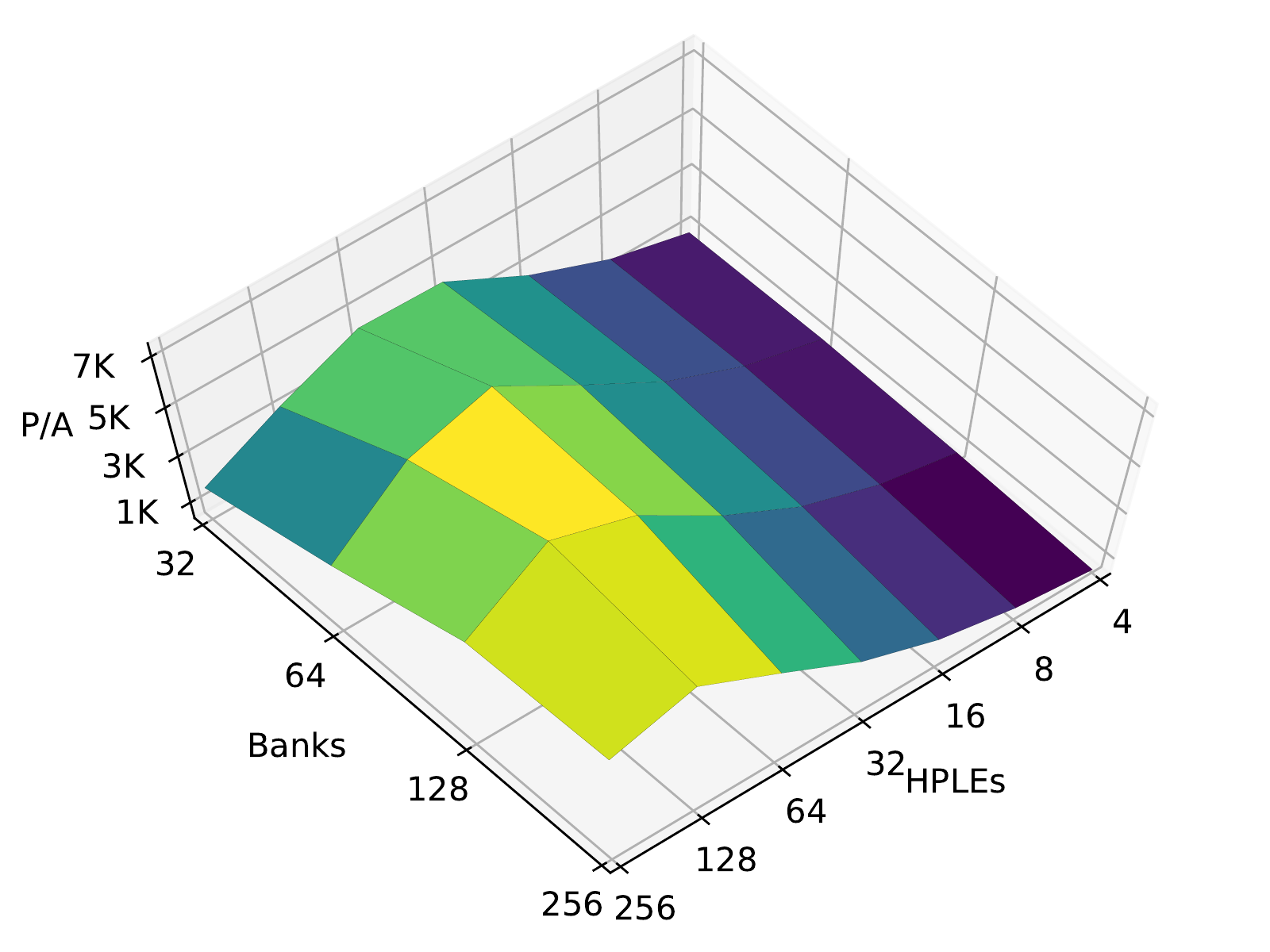}
  \vspace{-1em}
  \caption{\bf Performance per area (P/A) of 64K NTT under different RPU configurations.}
  \label{fig:ntt_efficiency_contour}
  %\vspace{-1em}
\end{figure}
\par Figure~\ref{fig:ntt_efficiency_contour} shows the performance per area (P/A) of different RPU configurations running 64K NTT.
High P/A indicates that the design achieves faster execution per $mm^{2}$.
In Figure~\ref{fig:ntt_efficiency_contour}, the lower P/A values are indicated by dark blue and higher P/A values are depicted using light yellow. 
The (128, 128) RPU configuration is the most efficient configuration, and (64, 64) has a second-best efficiency. 
With 128 HPLEs, we observe P/A improvement up to 128 banks and P/A drops for 256 banks. 
This is because further increasing banks provides no speedup while significantly increasing the vector crossbar area by 2$\times$, which is the principal contributor to the efficiency drop.
For 128 VDM banks, we observe P/A improvement up to 128 HPLEs and P/A drops for 256 HPLEs. 
This is because at (256, 128), performance improvement is only 16\% compared to (128, 128) while the large HPLE area doubles.
With 256 HPLEs, only two vector elements per instruction are mapped to an HPLE, which can create bubbles as the front-end cannot find the next compute instruction fast enough.
In practice, going beyond 128 HPLEs is not practical as crossbar cost begins to dominate at this point (see Section~\ref{subsec:area_profling}).
\subsection{\bf Area Breakdown}
\label{subsec:area_profling}
\begin{figure*}[ht!]
\centering
\hspace{-0.52in}
\subfigure[]{
\includegraphics[width=0.385\textwidth]{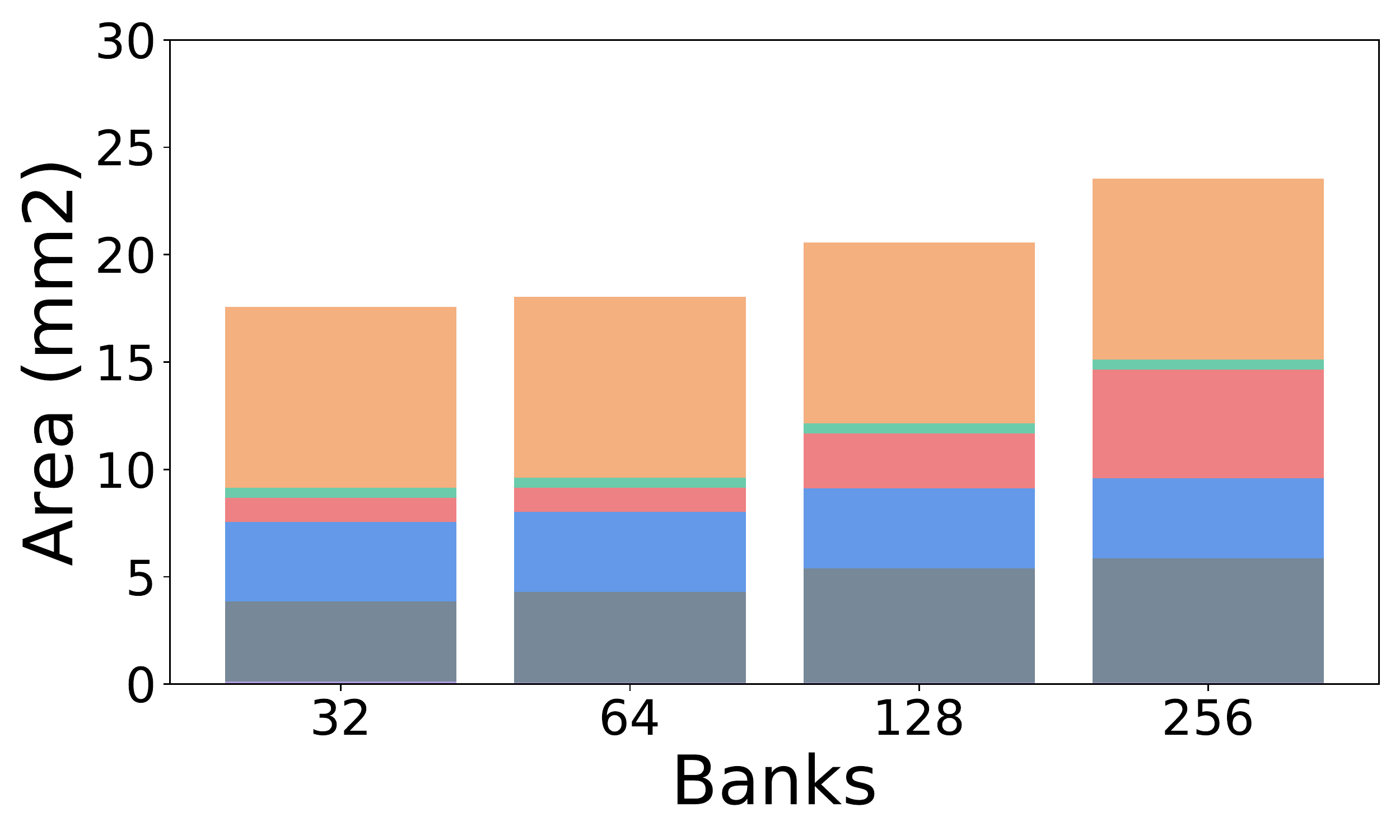}
\label{fig:ntt_bank_profiling}	
}
\hspace{-0.2in}
\subfigure[]{
\includegraphics[width=0.385\textwidth]{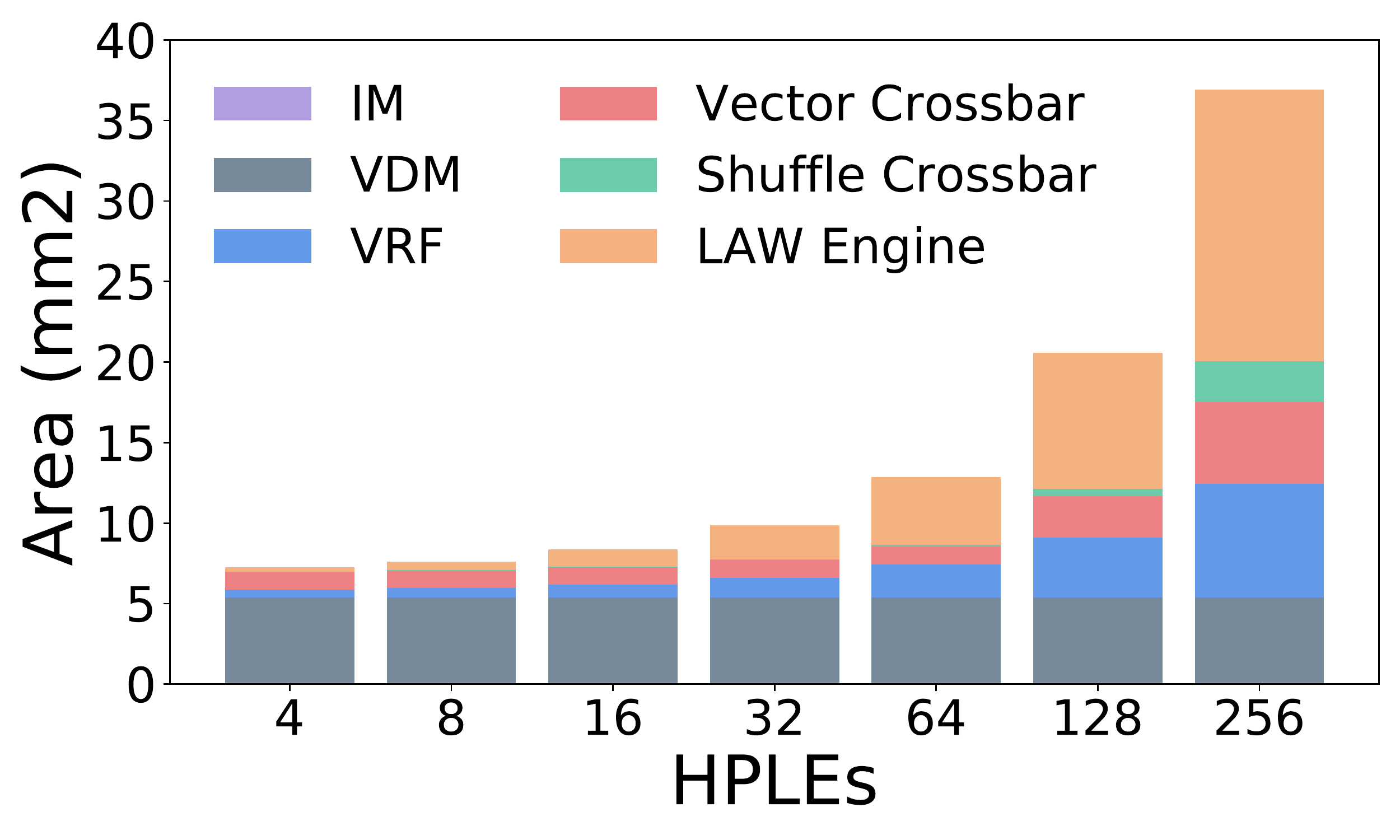}
\label{fig:ntt_hple_profiling}	
}
\hspace{-0.2in}
\subfigure[]{
\includegraphics[width=0.25\textwidth]{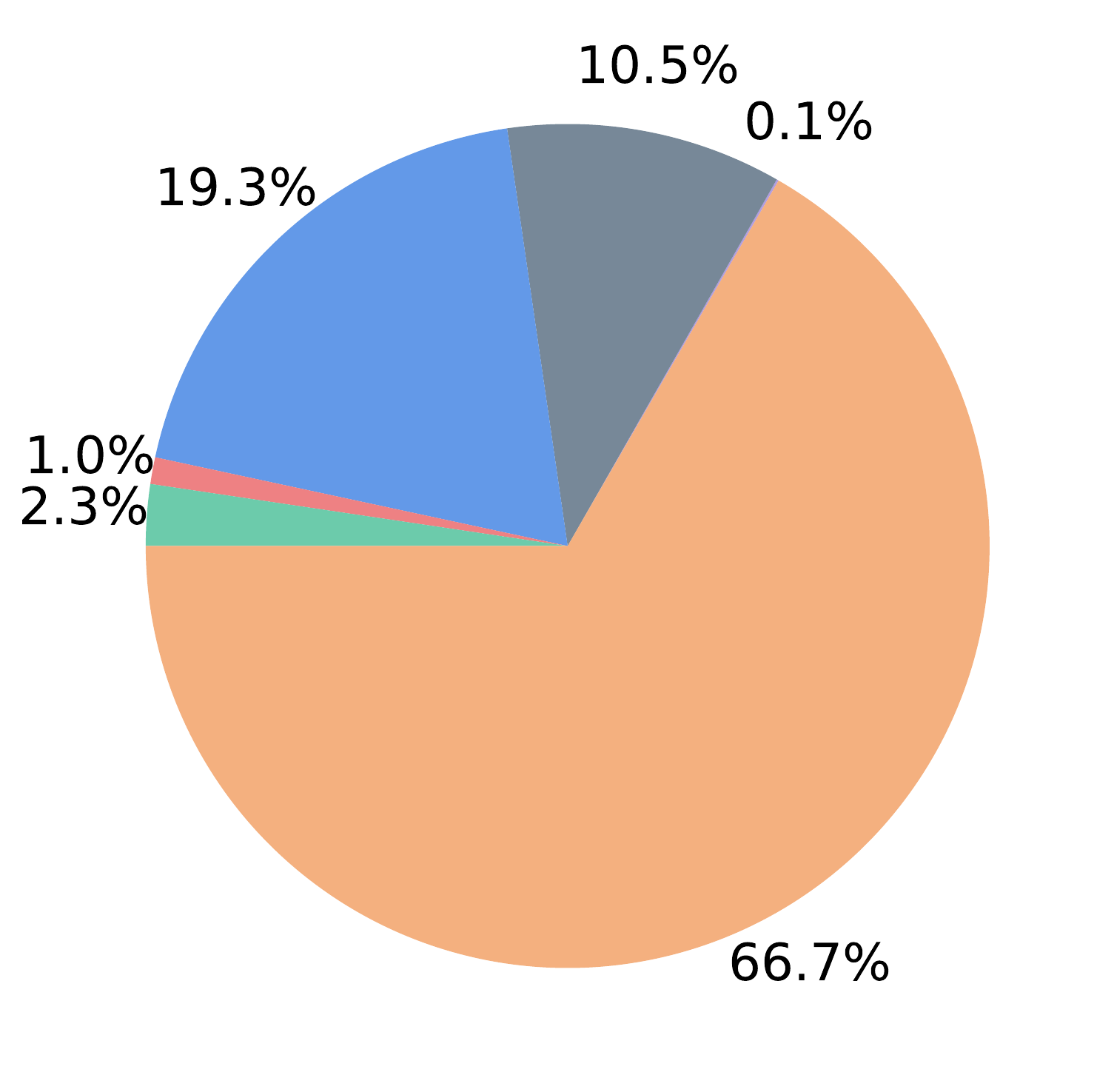}
  \label{fig:ntt_energy_pie}
}
\hspace{-0.52in}
\vspace{-0.5em}
\caption{\bf (a) RPU area breakdown, sweeping VDM banking and fixing 128 HPLEs and (b) sweeping HPLEs and fixing 128 VDM banks. (c) 64k NTT energy breakdown on a (128, 128) RPU.}
\vspace{-1.5em}
\label{fig:ntt_area_energy}	
\end{figure*}
Figure~\ref{fig:ntt_bank_profiling} and Figure~\ref{fig:ntt_hple_profiling} show the area breakdown of RPU components: Instruction Memory (IM), Vector Data Memory (VDM), High performance LAW Engine (HPLE) (constituting a Vector Register File (VRF) slice and LAW Engine), Vector Crossbar (VBAR), and Shuffle Crossbar (SBAR) when varying the number of HPLEs and VDM banks.
Figure~\ref{fig:ntt_bank_profiling} shows an area breakdown of RPU components when the number of HPLEs is fixed and we increase VDM banks. 
As expected, we observe that VBAR scales poorly as the number of banks increases, because the VBAR needs to accommodate parallel transfer between VDM banks and HPLEs. For 128 HPLEs, we find that the VBAR area remains minimal for up to 64 VDM banks. However, beyond this point, the VBAR area doubles when doubling the number of VDM banks.
Banking the VDM has less of an impact on the overall area. As the VDM banks double, RPU area increases by 10\%-24\%.
The remaining components are not changed.
\par In Figure~\ref{fig:ntt_hple_profiling}, we fix VDM banks to 128 and sweep HPLEs.
Here, we find that as the number of HPLEs doubles, the area of LAW Engine also doubles, while the area of the VRF jumps by 1.5$\times$-2$\times$.
The VRF is divided between the HPLEs, and each VRF slice (512 divided by HPLEs) has 64 registers. Therefore, by increasing the number of HPLEs, the VRF slice gets smaller.
Smaller VRF slice is mapped using small but inefficient memory macros, which stores fewer bits per $mm^{2}$. 
E.g., 512B single-port memory, has 2010$\mu m^2$ area, storing 255 KB/$mm^{2}$, while 256B memory, has 1818$\mu m^{2}$ area, storing 140 KB/$mm^{2}$. 
VBAR area remains minimal and more than doubles with doubling the number of HPLEs. As the number of HPLEs doubles, the SBAR area triples. However, for 256 HPLEs, the SBAR area is $5\times$ larger compared to 128 HPLEs.
\subsection{\bf Energy Consumption}
Figure~\ref{fig:ntt_energy_pie} shows the energy breakdown of 64K NTT on a RPU with 128 HPLEs and 128 VDM banks. 
The total energy required for 64K NTT execution is 49.18$\mu J$. 
The LAW Engine and VRF consume 66.7\% and 19.3\% of the total energy, respectively, combining for 86\% of the total. 
The VDM, which has 8$\times$ the memory size of the VRF, consumes only 10.5\% of the energy, as the VDM employs large and efficient memory macros and is accessed less than VRF. 
We also find that the 128b modular multipliers dissipate significant power at 104mW each; in the future we plan to research new low-power designs.
The vector crossbar and shuffle crossbar consume 2.3\% and 1.0\% of the total energy, respectively. 
The total average power of the (128, 128) RPU is 7.44$W$.
\subsection{\bf Impact of Code Optimization}
\label{subsec:code_opt}
\begin{figure}[t!]
  \centering
\includegraphics[width=0.45\textwidth]{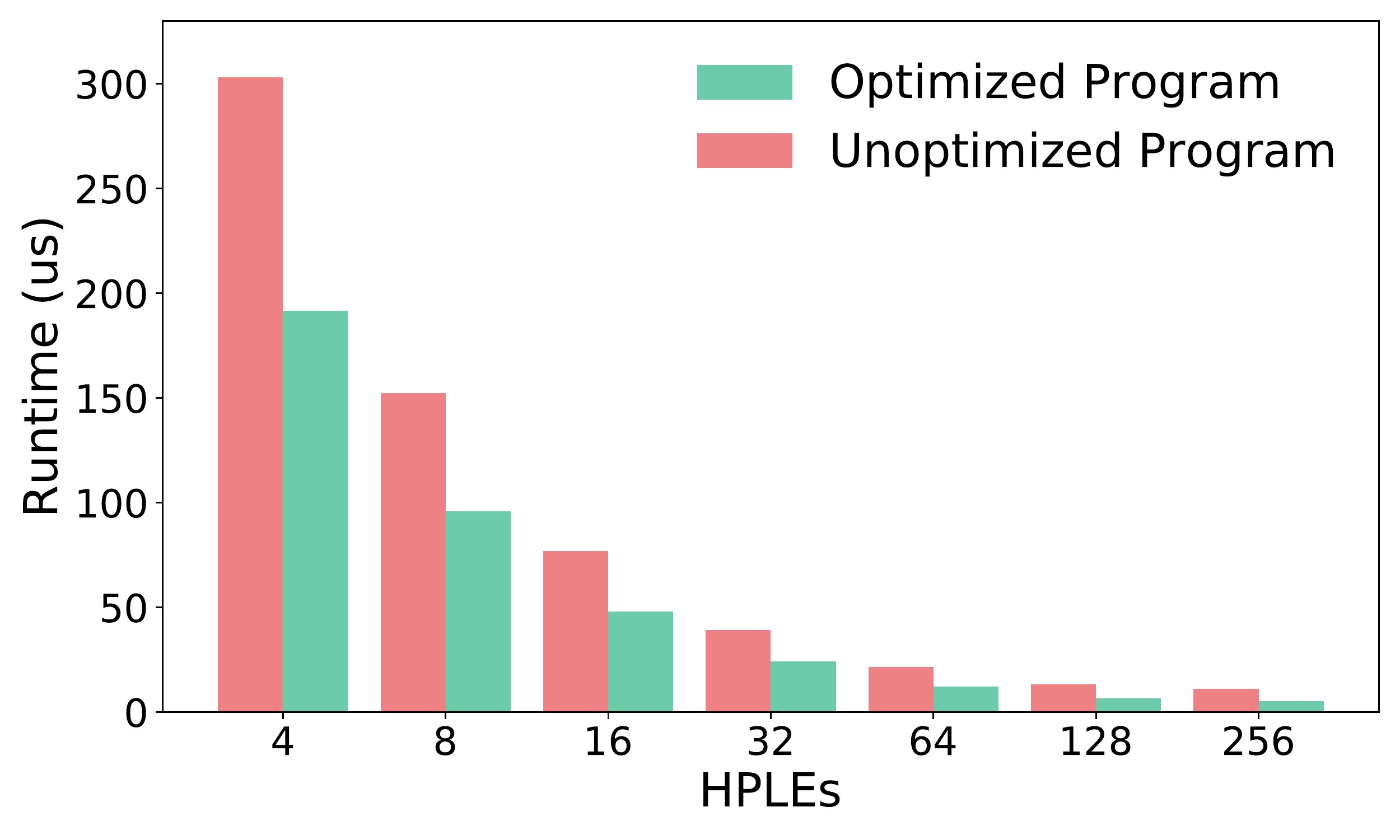}
\vspace{-1.5em}
\caption{\bf 64K NTT runtime for optimized and unoptimized code.}
\label{fig:code_compare}
\vspace{-1.5em}
\end{figure}
Figure~\ref{fig:code_compare} shows the runtime of two programs for executing a 64K NTT, one naive and one optimized with SPIRAL. 
We vary the number of HPLEs and keep the number of banks at 128.
The unoptimized program 
has no knowledge of the RPU micro-architecture. 
The optimized program is aware of the design and schedules instructions for maximum parallelism, as described in Section~\ref{sec:SPIRAL}. 
Figure~\ref{fig:code_compare} shows that a program considering the underlying microarchitecture of RPU is on average 1.8$\times$ faster than an unoptimized program. Hardware-aware optimized program effectively schedules code compared to unoptimized program. For example, when using 256 HPLEs, shuffle instructions in unoptimized code wait for 3,840 clock cycle to release resources from busyboard. 
On the other hand, shuffle instruction in optimized code wait only 128 clock cycles.
This is because in the unoptimized program, the shuffle, like other instructions, is always stalled waiting for the result of the previous instruction, whereas SPIRAL intersperses independent instructions.
\subsection{\bf RPU Sensitivity}
\begin{figure}[t!]
  \centering
\includegraphics[width=0.4\textwidth]{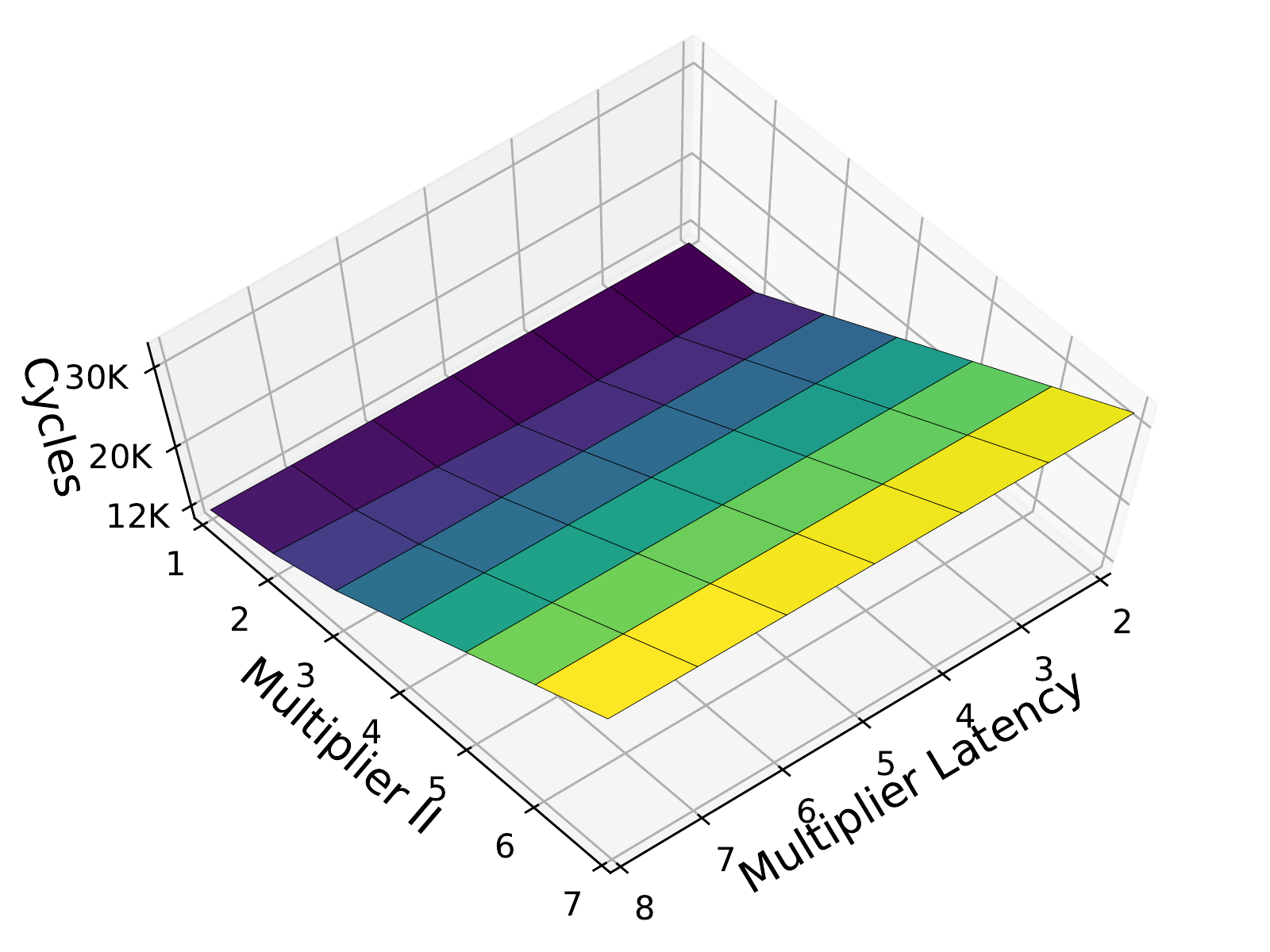}
  \vspace{-1em}
\caption{\bf  RPU sensitivity towards multiplier latency and II while running 64K NTT for (128, 128) configuration.}
\label{fig:mult_sense}
  \vspace{-1em}
\end{figure}

We now conduct performance sensitivity studies of the RPU.
Figure~\ref{fig:mult_sense} shows the clock cycles needed to compute a 64K NTT as a function of multiplier pipeline depth (latency) and initiation interval (II).
We first observe that the RPU is not highly sensitive to multiplier latency, which is intuitive as all units are fully pipelined.
However, clock cycles generally increase by 1.5$\times$ when we increase II.
This is natural as increasing II decreases the throughput of the functional units. 
Interestingly, we find an II of 2 only increases clock cycles by 16\%. 
This is because the 64K NTT has 1024 CIs and 1920 SIs and SIs create bottleneck.

\begin{figure}[t!]
  \centering
\includegraphics[width=0.4\textwidth]{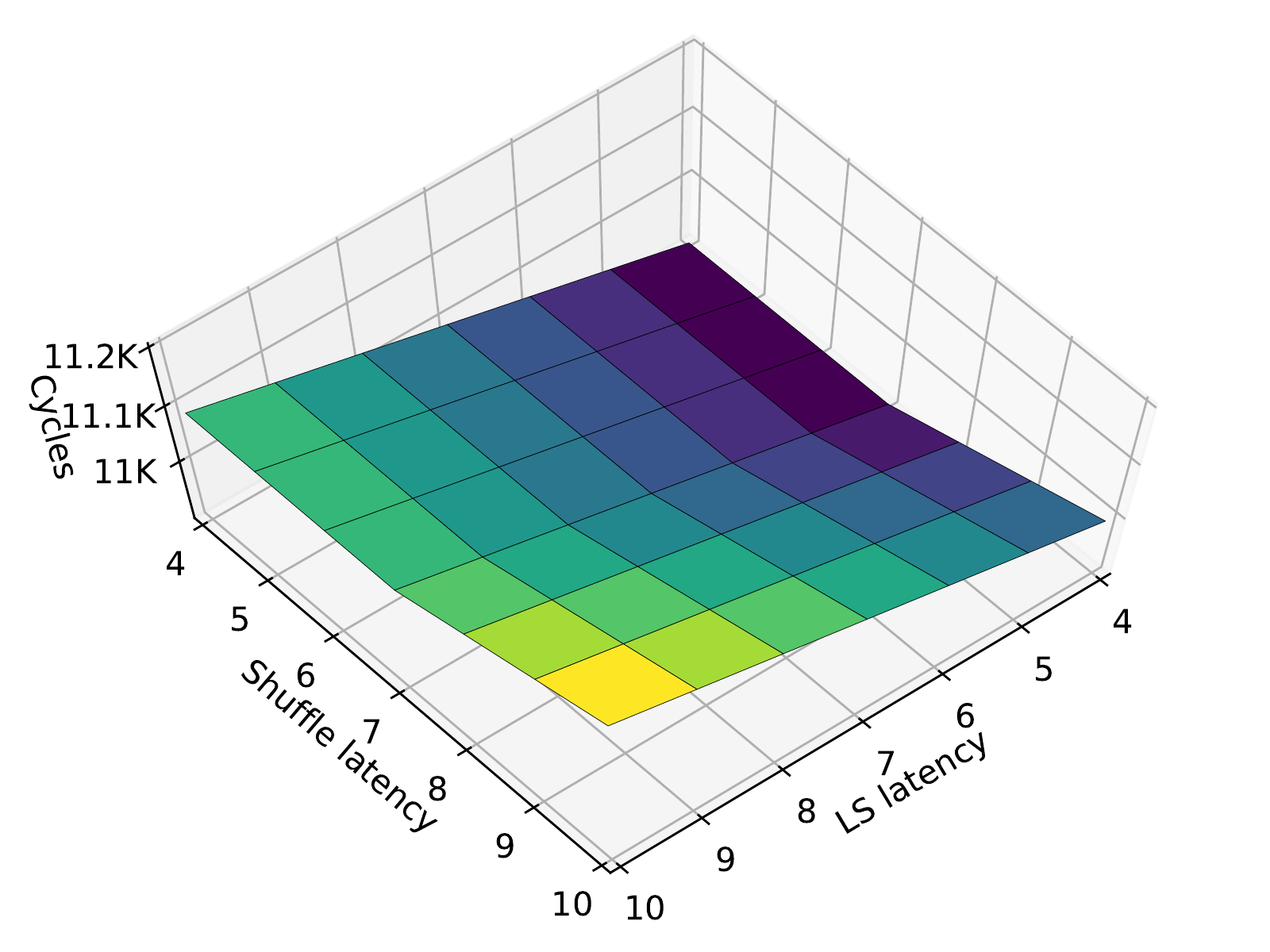}
%\vspace{-1em}
\caption{\bf RPU sensitivity towards shuffle crossbar and vector crossbar while running 64K NTT for (128, 128) configuration}
\label{fig:crossbar_sense}
%  \vspace{-4mm}
\end{figure}
\par Figure~\ref{fig:crossbar_sense} shows the clock cycle required for 64K NTT code when changing the SBAR (shuffle latency) and VBAR (LS latency).
Figure~\ref{fig:crossbar_sense} shows that total cycles increase slightly when increasing LS latency. 
If shuffle latency is constant, a LS latency of 10 has 1.7\% clock cycle increment compared to LS latency of 4.
If LS latency is constant, clock cycles do not change if we increase shuffle latency to 7. 
For higher shuffle latency, the total cycle count increases marginally with shuffle latency. 
\par There are two key takeaways from the sensitivity study: 
(1) This study guides us in hardware component selection.
For example, we can select a small multiplier with II=2 for the LAW engine as a larger multiplier with II=1 has a similar runtime. We can also increase shuffle latency if the area is reduced.
(2) RPU sensitivity study can help optimize code. 
For example, even though NTT has fewer load/store instructions compared to shuffle instructions, the RPU is more sensitive to load/store latency.  
Therefore, we can focus more on scheduling load/store instructions for further optimization.
\subsection{\bf Performance Comparison}
\begin{figure}[t!]
  \centering
\includegraphics[width=0.45\textwidth]{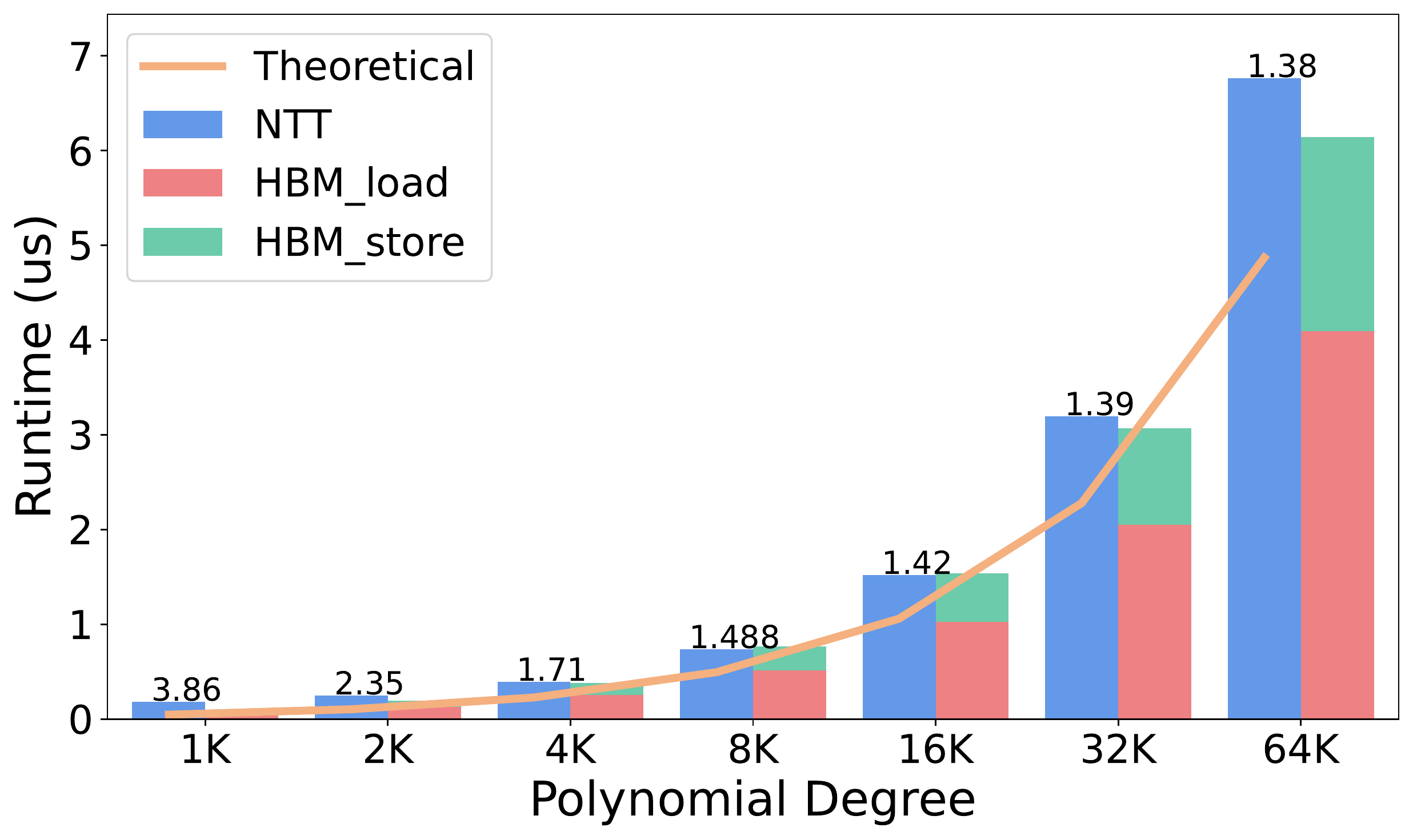}
\caption{\bf Comparing NTT runtime on RPU with the theoretical latency and load store latency for HBM2.}\label{fig:polysize}
\end{figure}
The VDM interacts with a high-bandwidth off-chip memory, HBM2. We assume a 512GB/s HBM2 as in prior works~\cite{choquette2021nvidia, feldmann_f1_2021}.
We have evaluated the runtime of NTT for polynomial degrees between 1K to 64K using the (128, 128) RPU.
Figure~\ref{fig:polysize} compares the obtained results for RPU with HBM2 and the theoretical/ideal case, explained below. 
\par HBM load and store show the time it takes to load data from off-chip memory (HBM2) to VDM and store back the NTT result. 
The theoretical latency is the ideal time it would take to finish NTT's execution using the allotted multipliers, ignoring data movement and data dependence.
NTT includes $\mathcal{O}(n\log n)$ compute instructions, which in our design are divided between HPLEs all working in parallel. 
We calculate the theoretical latency using $\frac{n\log_2 n}{HPLEs * frequency}$, where $n$ is the polynomial degree. 
Labels on each bar represent the ratio of NTT's runtime on RPU over the theoretical runtime. 
Figure~\ref{fig:polysize} shows how as the polynomial degree increases, the ratio decreases.
For 1K, the RPU is $3.86\times$ slower than theoretical runtime, while it reduces to $1.38\times$ for 64K. 
For larger polynomial degrees, there is more parallel work to do on the RPU, making the RPU's runtime closer to the theoretical limit. 
\par To overlap the execution of NTT with loading data from HBM2, HBM2's latency should be less than or equal to NTT runtime.
For 64K to 8K, the runtime drops more than two times as we half the polynomial degree, e.g., 32 NTT is $2\times$ faster than 64K. 
However, HBM2’s latency is proportional to the polynomial degree. As a result, for 16K, HBM2 becomes 0.041us faster than NTT, which is only 2.77\% of 16K NTT's runtime. 
For NTTs smaller than 8K, the amount of work per loaded data decreases, as we half the polynomial degree the runtime drops less than $2\times$. 
This trend causes the HBM2 latency to again become less than NTT. In total, a 512GB/s HBM2 satisfies the off-chip bandwidth requirement for NTT execution on RPU for various polynomial degrees. 
\par
\begin{figure}[t!]
  \centering
\includegraphics[width=0.45\textwidth]{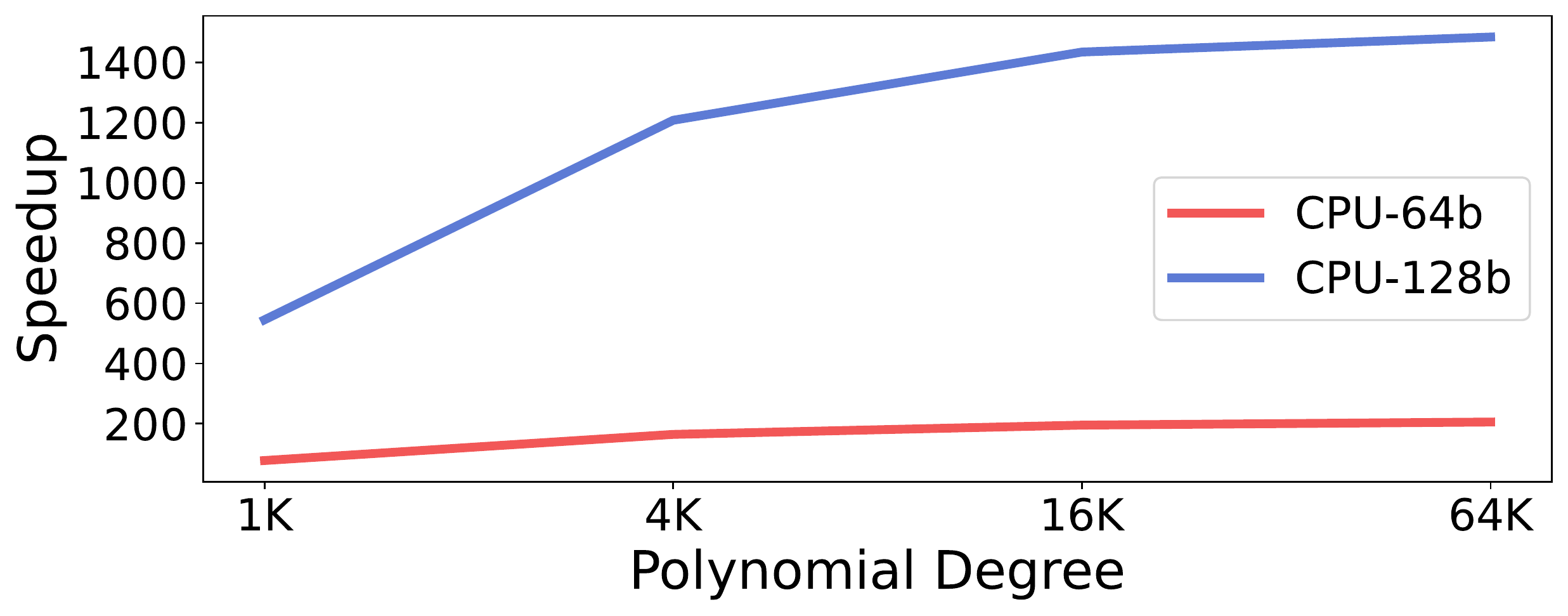}
\caption{\bf RPU speedup for various polynomial degrees over CPU for 64bit and 128bit data.}\label{fig:cpu}
  \vspace{-.5em}
\end{figure}

 Figure \ref{fig:cpu} illustrates the RPU's speedup over the CPU. 
We use a 32-core 2.5GHz AMD EPYC 7502 and run different polynomial degrees of NTT provided by OpenFHE benchmarks \cite{OpenFHE} for 64-bit and 128-bit data on the CPU. Our designed RPU achieves a speedup of $545\times$ to $1484\times$ over CPU implementation of NTT for 128 bit data. Even if we run 64-bit data on the CPU (and use the 128-bit RPU), the RPU is still $77\times$ to $205\times$ faster. 
Therefore, by having a 128-bit design, we can support a wide range of data bitwidths and still get a high speedup over the CPU.
For polynomials larger than 8k, as we double the polynomial degree, the runtime increases more than $2\times$, as the data cannot fit in the VRF and the workload would require more load/stores from VRF to VDM. As a result, the slope of the speedup over CPU will start to decrease, reaching $1484\times$ for 64K. 

\section{\bf Related work}
Most prior work focuses on accelerating FHE primitives~\cite{ feldmann_f1_2021, reagen2021cheetah, riazi2020heax, kim2022bts, samardzic2022craterlake, nabeel2022cofhee}. 
F1 \cite{feldmann_f1_2021} designs specialized functional units to accelerate the primitive computations shared among various FHE schemes. 
One of those specialized functional units is designed to accelerate NTT. 
For a fair comparison between our RPU and F1, we consider NTTs' functional unit and vector register file area only for F1 and the HPLE and VRF for RPU. 
F1 uses 32-bit data; therefore, we scale their reported area by 4, as a conservative assumption given multiplier scale quadratically with input size, to match the RPU's 128-bit design. 
In addition, because we do not have a multi-RPU design, we only assume one compute cluster for F1. 
Considering these assumptions, a 16K NTT on F1 would take 2864ns to execute, with an area of 11.32 $mm^{2}$. 
The RPU takes 1500ns, while the area is 12.61 $mm^{2}$. F1's throughput/area is $2\times$ more than RPU. However, F1 only supports up to 16K polynomial degrees, while our work has no limit.
Also note that in this comparison, F1 can only process NTTs, as we did not count the area of the other functional units.
Additionally, F1 uses 32b data, while we use 128b data as it has been shown to be needed for 128b security without any information leakage\cite{li2022securing}. 
To conclude, our RPU's throughput/area is close to the fixed approach used in F1 while the RPU is not limited in the polynomial degree and flexible for different existing and future ring-based applications, which may require higher polynomial degrees.
\par Comparing with GPU, according to \cite{ozerk2022efficient}, a 64K, 30-bit NTT an a V100 is 6x slower than the RPU using 128-bit. In addition, the RPU uses $40\times$ less area and $40\times$ less power than a V100. If 128-bit is used on a GPU, we expect the RPU speedup to improve significantly.
\par Recently, several HE compilers have been developed. Porcupine\cite{cowan2021porcupine} generates optimized and vectorized HE kernels through program synthesis. Ramparts \cite{archer2019ramparts}, directly translates Julia functions to HE workloads using PALISADE\cite{palisade_new} library. nGraph\cite{boemer2019ngraph} has added support for CKKS and BGV schemes to an existing machine learning compiler. Eva\cite{dathathri2020eva} introduces a new input language for vector arithmetic that targets the SEAL library and hides the cryptographic complexities from the programming.
\section{Conclusion}

This paper develops the RPU to accelerate RLWE-based workloads.
The RPU is an accelerator that realizes our proposed ISA, B512.
In addition to the hardware, we propose a compiler flow based on SPIRAL to automate high-performance programming.
Using a newly developed simulator, we rigorously explore the design-space to understand tradeoffs and identify efficient designs.
We show that an RPU with 128 banks and HPLEs, can execute 128-bit 64K NTT in 6.7 $\mu$s using 20.5mm$^2$ of GF 12nm, achieving a speedup of 1485$\times$ over a CPU. 

\section*{Acknowledgment}
This work was supported in part by the Applications Driving Architectures (ADA) Research Center, a JUMP Center co-sponsored by SRC and DARPA. Additionally, this research was developed with funding from the Defense Advanced Research Projects Agency (DARPA), under the Data Protection in Virtual Environments (DPRIVE) program, contract HR0011-21-9-0003. 
Reagen received supported from the NY State Center for Advanced Technology in Telecommunications (CATT).
The views, opinions, and/or findings expressed are those of the authors and do not necessarily reflect the views of the sponsors.
\bibliographystyle{ieeetr}
\bibliography{main}

\end{document}